\documentstyle[twoside, epsf, hyperref]{article}

\input ibvs3.sty

\begin{document}

\newcommand{\Rsun}{\hbox{$R_{\odot}$}}
\newcommand{\Msun}{\hbox{$M_{\odot}$}}
\newcommand{\A}[1]{\hbox{$a_{\rm #1}$}}
\newcommand{\K}[1]{\hbox{$K_{\rm #1}$}}
\newcommand{\LL}[1]{\hbox{$L_{\rm #1}$}}
\newcommand{\M}[1]{\hbox{$M_{\rm #1}$}}
\newcommand{\R}[1]{\hbox{$R_{\rm #1}$}}
\newcommand{\V}[1]{\hbox{$V_{\rm #1}$}}
\newcommand{\Ome}[1]{\ensuremath{\Omega_{\rm #1}}}
\newcommand{\Teff}{\hbox{$T_{\rm eff}$}}
\newcommand{\Tmult}{\hbox{$T_{\rm mult}$}}
\newcommand{\T}[1]{\hbox{$T_{\rm #1}$}}
\newcommand{\x}[1]{\hbox{$x_{\rm #1}$}}
\newcommand{\Abol}{\hbox{$A_{\rm bol}$}}
\newcommand{\Tspot}{\hbox{$T_{\rm spot}$}}
\newcommand{\rspot}{\hbox{$r_{\rm spot}$}}
\newcommand{\RV}[1]{\hbox{RV$_{\rm #1}$}}
\newcommand{\rside}[1]{\hbox{$r_{\rm side,#1}$}}
\newcommand{\rpole}[1]{\hbox{$r_{\rm pole,#1}$}}
\newcommand{\geff}{\hbox{$g_{\rm eff}$}}
\newcommand{\vsini}{\ensuremath{v\hspace{0.005in}\sin\hspace{0.005in}i}}
\newcommand{\resid}{\ensuremath{\mid\ell_{\rm obs}-\ell_{\rm cal}\mid_{_{\rm ave}}}}
\newcommand{\ibvs}{IAU Inf.\ Bull.\ Var.\ Stars}
\newcommand{\obs}{Obs.}
\newcommand{\kmps}{km~s$^{-1}$}
\newcommand{\SAMEC}{\hbox{\rm\small SAMEC}}
\newcommand{\chisq}{\hbox{$\chi^{\rm 2}$}}

\IBVSheadDOI{63}{6262}{27 March 2019}

\IBVStitletl{On the Period and Light Curve of the}{A-Type W~UMa binary GSC~3208~1986}
 
\begin{center}
\IBVSauth{EATON, JOEL A.$^1$; ODELL, ANDREW P.$^2$; POLAKIS, THOMAS A.$^3$}
\end{center}

\IBVSinst{7050 Bakerville Road, Waverly, TN 37185 USA; e-mail: eatonjoel @ yahoo.com}
\IBVSinst{Dept of Physics and Astronomy, NAU Box 6010, Flagstaff AZ 86011 USA; e-mail: WCorvi @ yahoo.com}
\IBVSinst{Command Module Observatory, 121 W. Alameda Dr., Tempe, AZ 85282 USA; e-mail: tpolakis@cox.net}

\SIMBADobj{GSC 3208 1986}

\IBVStyp{W UMa}
\IBVSkey{Binaries: eclipsing}

\IBVSabs{We present a new period study and light-curve solutions for the A-Type W UMa binary GSC 3208 1986.  Contrary to a previous claim by R.G. Samec et al. of a rapidly decreasing period, the system's period is $increasing$ moderately on a timescale of 2E6 years.  The light curve is variable on the time scale of years, which can be understood by changes in how much it overfills its Roche lobe.}

\begintext

Contact binaries are binaries close enough that their components are enclosed in a common, 
probably convective envelope (Lucy 1968).  The best known members of this class are the 
W~Ursae~Majoris systems (Binnendijk 1970), although there are other rarer binaries that 
may be in marginal contact (e.g., Ka{\l}u\.{z}ny 1983, 1986a--d; Siwak et al.\ 2010).  
Binnendijk (pp.\ 218-221) defined two varieties of these W UMa systems, A-types, with transit 
primary eclipses, and W-types, with occultation primaries.  Given the direct dependence of 
the ratio of radii on mass ratio in contact binaries, these A- and W-type classes correspond 
to $q$=\M2/\M1 less than and greater than 1.0, respectively.

GSC 3308 1986 ($\alpha$(2000)=22\hr 25\mm 16\fsec0, $\delta$(2000)=+41\deg 27\arcm 51\farcs9) 
is a faint A-type W~UMa binary observed and analyzed by Samec et al.\ (2015a; hereafter \SAMEC).  
\SAMEC\ obtained four nights of photometry ($\sigma_{\rm B}$$\approx$0.006) and found an F3~V 
spectral type from a spectrum taken at the Dominion Astrophysical Observatory, a mass ratio 
$q$=0.24, and that the star overfills its Roche lobe by 39\%.  These properties are not surprising 
for such a system, but \SAMEC\ also derived a very rapid period $decrease$, corresponding to a 
timescale of 3$\times$10${^5}$ years.  This seems unlikely for what they claim is an ``ancient" 
contact system, especially if caused by magnetic braking, their favored period-change mechanism.

\paragraph{\bf EPHEMERIS:} Suspecting that the radical period decrease might result from 
R.\ G.\ Samec's previously documented (Odell et al.\ 2011) error of confusing Modified Julian 
Date (Heliocentric Julian Date - 2,400,000.5) with Reduced Julian Date (HJD - 2,400,000.0) 
in data from the Northern Sky Variability Survey (NSVS, see Wozniak et al.\ 2004), we obtained 
the archival data from the NSVS and SuperWASP (SWASP, see Butters et al.\ 2010) web sites.
We have subsequently obtained new light curves for 2017 and 2018 (Polakis; BVRI on the 
UBV/Cousins system; Table 1, given as file 6262-t1.txt on the IBVS web site) and added the 
published photometry of Liakos \& Niarchos (2011) and \SAMEC\ to give nine seasonal light curves.  
Using these, we find a very different result than \SAMEC.  We have derived new effective times 
of minimum for these nine epochs by fitting those seasonal light curves with the Wilson-Devinney 
code to measure phase shifts with respect to the ephemeris of Eq.\ 1.  These are listed in 
Table 2; the errors given are the $\sigma$'s calculated by the W-D code multiplied by a factor 
of three per Popper (1984).

\IBVSedata{6262-t1.txt}
\IBVSdataKey{6262-t1.txt}{GSC 3208 1986}{BVRI Data}

\begin{table}[!ht]
\small
\centerline{{\bf Table 2.} $O-C$ Residuals for linear and quadratic elements (days)}
\begin{center}
\begin{tabular}{lcccl}
\hline
\multicolumn{1}{c}{Epoch (Obs)}& Cycle  & (Obs-Calc) & (Obs-Calc)  & Source of data\\
\multicolumn{1}{c}{RJD}    &    (N)     &   linear   & quadratic   &            \\
                           &            &  (Eq.\ 1)  &  (Eq.\ 2)   &            \\
\hline
51464.1096 $\pm$ 0.0010 & -11693&   0.0022 & -0.0017   &  NSVS             \\
53247.4351 $\pm$ 0.0003 &  -7285&  -0.0005 &  0.0003   &  SWASP 2004       \\
53989.8134 $\pm$ 0.0006 &  -5450&  -0.0014 &  0.0003   &  SWASP 2006       \\
54324.7939 $\pm$ 0.00011&  -4622&  -0.0018 &  0.0000   &  SWASP 2007 Epoch1\\
54374.1509 $\pm$ 0.00013&  -4500&  -0.0019 & -0.0001   &  SWASP 2007 Epoch2\\
55410.2457 $\pm$ 0.0005 &  -1939&  -0.0013 & -0.0001   &  Liakos\&Niarchos \\
56194.7011 $\pm$ 0.0003 &      0&   0.0000 & -0.0001   &  Samec            \\
57925.8458 $\pm$ 0.0003 &   4279&   0.0055 & -0.0003   &  Polakis 2017     \\
58415.7787 $\pm$ 0.0002 &   5490&   0.0081 &  0.0001   &  Polakis 2018     \\
\hline
\end{tabular}
\end{center}
\end{table}

In analyzing the period, we first used a preliminary linear ephemeris derived by Odell from 
the NSVS plus Polakis 2017 data, namely 
\begin{equation}
  {\rm HJD\ T_{min}\ I} = 2,456,194.7011 + 0.4045663 \times {\rm N},
\end{equation}
to phase all the data into annual/seasonal light curves, then derived the deviations 
of the phases from this linear ephemeris with the W-D code as noted above, and then 
fit those deviations with a second-order polynomial to determine the following 
quadratic ephemeris:
\begin{equation}
  {\rm HJD\ T_{min}\ I} = 2,456,194.7012(1) + 0.40456718(1) \times {\rm N} + 1.03(5)\times10^{-10} \times {\rm N}^2.
\end{equation}
In this equation the numbers in parentheses are errors in the last decimal place, 
and N is the cycle number.  Fig.\ 1 shows the deviations from Eq.\ 1 and the 
quadratic fit.

\IBVSfig{7.5cm}{6262-f1.ps}{(O-C) Diagram for GSC 3208 1986.}
\IBVSfigKey{6262-f1.ps}{GSC 3208 1986}{(O-C) Diagram}


\paragraph{\bf SPECTRA:} Odell obtained two spectra of GSC~3208~1986 with the Boller\&Chivens 
Spectrograph on the Steward Observatory 90-inch telescope around 1 June 2015, specifically at 
HJD~2,457,173.9734 (phase 0.55) and HJD~2,457,174.8694 (phase 0.76).  These spectra covered 
the wavelength range 3900--4750~\AA\ and are consistent with the F3~V spectral type of \SAMEC.  
They give radial velocities for the components of \RV1=22.1$\pm$7.2 \kmps\ for the phase near 
conjunction and \RV1=86.9$\pm$8.2 \kmps\ and \RV2=-298$\pm$25 \kmps\ for the quadrature.  These 
values give a crude indication of the velocity amplitudes of the components, \K1=91$\pm$16 \kmps\ 
and \K2=294$\pm$25 \kmps\ with $\gamma$=-4 \kmps.  The resulting spectroscopic mass ratio 
$q$=0.30$\pm$0.03 is $\sim$ consistent with the photometric mass ratio.

\paragraph{\bf LIGHT CURVE:} The extensive observations from SWASP give us the opportunity to solve 
well-defined light curves for the three years, 2007, 2006, and 2004.  The data for 2007 are by far 
the best and most numerous, so we will concentrate on them.  Consequently, we have formed 200 normal 
points derived from the roughly 11,300 SWASP observations for 2007, giving them in Table 3 
(available through the IBVS website as {\tt 6262-t3.txt}) as orbital phase (based on Eq.\ 1), 
magnitude, and a standard deviation of the mean for each magnitude.  The typical normal point 
has an uncertainty of $\sigma$=0.0019 mag (S.D.), nominally giving about the same total weight as 
the photometry published by \SAMEC, but the SWASP data cover enough time to average out the typical 
wavelength-independent observational errors of data taken on a mere four nights.  These data 
represent a broad band in the optical, corresponding roughly to $V$ of the $UBV$ system.  Fig.\ 2 
shows the SWASP light curves for 2007 (Table 3) with a representation of the solution of Table 4 
plotted as a solid line.

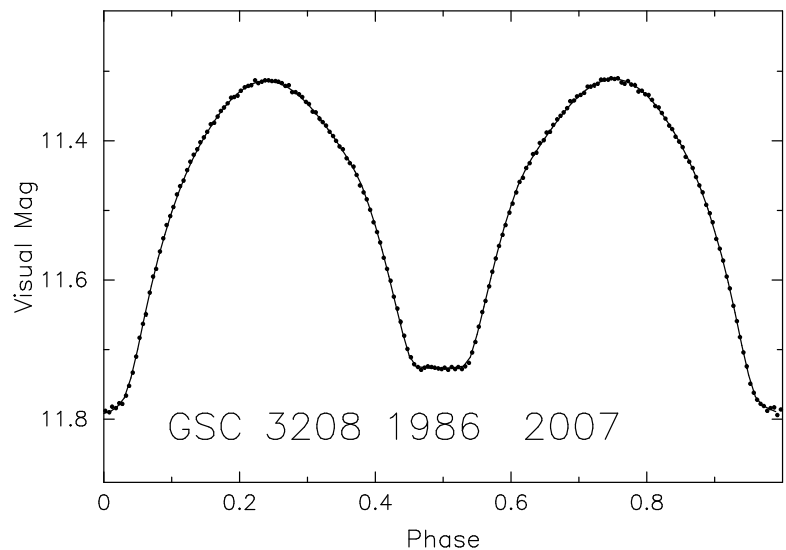
\IBVSfig{7.5cm}{6262-f2.ps}{Light-Curve Solution for SWASP, normal points for 2007.}
\IBVSfigKey{6262-f2.ps}{GSC 3208 1986}{Light-Curve Solution}

\IBVSedata{6262-t3.txt}
\IBVSdataKey{6262-t3.txt}{GSC 3208 1986}{normal points}

\vspace*{-0.5cm}

We have solved this light curve with the Wilson-Devinney code [2003 version; see Wilson \& Devinney 
(1971); Wilson (1990,94)], finding the elements in the second column of Table 4.  These are roughly 
consistent with \SAMEC's solution (Table 4, Col.\ 4).  In calculating this solution we adopted 
\SAMEC's temperature of the primary, convective gravity darkening (Lucy 1967), convective reflection 
effect (Rucinski 1969), the Kurucz-atmospheres option in the W-D code, and a linear limb-darkening 
coefficient from Van Hamme (1993).  We accounted for a slight O'Connell effect in the normal points 
with a small dark spot on the leading hemisphere of the primary component.  The small \chisq\ indicates 
the model fits the data as well as can be expected.  For completeness, we calculated a solution for 
2007 with radiative gravity darkening and reflection effect, because in the past there was some inkling 
that these hotter A-type systems might be radiative, but the fit was much worse, by a factor of two 
in \chisq.  This radiative solution had a significantly lower fillout, 13\%, as expected from the 
well-known correlation between fillout and gravity darkening.

The other two years of SWASP data had somewhat different light curves which we have solved 
by varying those elements of the 2007 solution that might conceivably change on the timescale 
of a few years.  Some elements, such as $q$ and $i$, cannot change materially on such a short 
timescale, so we are left with temperatures and fillout that might change.  Keeping $q$, $i$, 
\T1 fixed, we get the solution in Col.\ 3 of Table 4 for 2004. A greater depth of both eclipses 
in 2004 led to a larger overfilling of the Roche lobe.  The solution for 2006 had a marginally 
larger fillout, 39\%, for the worst data of the three years ($\sigma$=0.014 mag).  The differences 
between 2007 and 2004 might conceivably result from a change in the photometric band of the 
observations, but it would require a shift at least as great as from $V$ to $B$ between the two 
years.  A shift of this magnitude is rather unlikely (see Butters et al.\ 2010, Fig.\ 1).

All of these solutions imply that the standard overcontact model fits GSC~3208~1986 well.  Values 
of \Tmult, which measures the ratio of \T2 as measured in W-D, Mode 3, to its value for W-D, Mode 1,
(no break in temperature at the neck between the components), are 1.0 for all practical purposes, 
so the temperature varies smoothly over the  surface as determined by the gravity-darkening law.  
The solution for a radiative envelope, however, does not have this property and gives a significantly 
worse fit, so the envelope is not likely to be radiative.


\begin{table}[!ht]
\small
\centerline{{\bf Table 4.} GSC 3208 1986: Light-Curve Solutions}
\begin{center}
\begin{tabular}{lccccc}
\hline
Parameter     &      2007-SWASP      &    2004-SWASP      &  2012-\SAMEC\       &   2017-Polakis   &   2018-Polakis   \\
~~~~(1)       &         (2)          &       (3)          &     (4)             &      (5)         &      (6)         \\
\hline
\\
\x1=\x2 (fixed)&    0.51             &   0.51             &  Non-linear       &0.63,0.51,0.41,0.33&0.63,0.51,0.41,0.33\\
$g$ (fixed)   &     0.32             &   0.32             &  0.32               &   0.32           &   0.32           \\
\Abol (fixed) &     0.50             &   0.50             &  0.50               &   0.50           &   0.50           \\
\\
$i$ (deg)     &  85.60 $\pm$ 0.27    &  85.60 (fixed)     &  85.8 $\pm$ 0.1     &   85.60 (fixed)  &   85.60 (fixed)  \\
$q$ (\M2/\M1) &  0.2424 $\pm$ 0.0011 &  0.2424 (fixed)    &  0.2374 $\pm$ 0.0002&   0.2424 (fixed) &   0.2424 (fixed) \\
$\Omega$      &  2.2811 $\pm$ 0.0020 &  2.269 $\pm$ 0.0020&  2.261 $\pm$ 0.001  &2.273 $\pm$ 0.0018&2.279 $\pm$ 0.0016\\
fillout       &   35.3$\pm$1.3\%     &   49.1$\pm$1.3\%   &   39$\pm$0.7\%      &   40.3$\pm$1.2\% &   36.8$\pm$1.0\% \\
\T1 (K, fixed)&   6875               &   6875             &   6875              &   6875           &   6875           \\
\T2 (K)       &   6757 $\pm$ 22      &   6789 $\pm$ 10    &   6760 $\pm$ ?      &   6745 $\pm$ 11  &   6725 $\pm$ 8   \\
\Tmult        &   0.9950$\pm$0.0032  &   1.0009$\pm$0.0014&   0.9968            &   0.9948         &   0.9909         \\
$\sigma$ (mag)&   0.0019/point       &   0.0066/point     &$\sim$0.006/point    &$\sim$0.013/point &$\sim$0.013/point \\
$\chi^2$/DOF  &       1.2            &       1.1          &  $\sim$1.44         &  $\sim$2.2       &  $\sim$1.0       \\
\\
&\multicolumn{4}{c}{Spot on the Primary Component}\\
\\
 lat,long (deg)&     0,270           &     0,270          &      none           &      none        &      none        \\
  \rspot (deg)&       1.7            &      1.7           &                     &                  &                  \\
  \Tspot      &     (black)          &    (black)         &                     &                  &                  \\
\hline
\end{tabular}
\end{center}
\end{table}

You may have noticed that the quoted errors of our solution for 2007 and \SAMEC's solution for 2012 
are inconsistent, although the two data sets have roughly the same weight 
(\hbox{\#points/$\sigma^2$}).  This probably results from the way such uncertainties are calculated.
If we calculate the uncertainty of each element independently of all the others, we get values
for the 2007 SWASP solution similar to those quoted by \SAMEC.  However, if we let elements 
$q$, $i$, $\Omega$, \T2, and the $x$'s vary simultaneously, we get the uncertainties listed.
Adding $g$ and \Abol\ to the mix gives even bigger uncertainties, doubling the reported uncertainty 
of $\Omega$.  This result confirms Popper's (1984) insinuation that the uncertainties derived by 
the W-D code are misleading.  It also points to the intuitive truth that our assumptions about 
limb darkening, gravity darkening, and reflection effect will inevitably bias the results for all 
these contact and near-contact binaries.  

\vspace*{0.4cm}

ACKNOWLEDGMENTS: We thank Steward Observatory for allocating the telescope time to obtain the spectra 
we used.  This paper makes use of data from the Data Release 1 of the WASP data (Butters et al.\ 2010) 
as provided by the WASP consortium, and the computing and storage facilities at the CERIT Scientific 
Cloud, reg. no.  CZ.1.05/3.2.00/08.0144, which is operated by Masaryk University, Czech Republic. 
It also uses data from the Northern Sky Variability Survey created jointly by the Los Alamos National 
Laboratory and University of Michigan.  

\vspace*{-0.3cm}

\references

Binnendijk, L., 1970,  \textit{ARA\&Ap}, \textbf{12}, 217  \BIBCODE{1970VA.....12..217B}  \DOI{ 10.1016/0083-6656(70)90041-3}

Butters, O. W. et al., 2010, \textit{A\&A} \textbf{520}, L10 (SuperWASP)  \BIBCODE{2010A&A...520L..10B}

Ka{\l}u\.{z}ny, J., 1983, \textit{AcA}, \textbf{33}, 345   \BIBCODE{1983AcA....33..345K}

Ka{\l}u\.{z}ny, J., 1986a, \textit{AcA}, \textbf{36}, 105   \BIBCODE{1986AcA....36..105K}

Ka{\l}u\.{z}ny, J., 1986b, \textit{AcA}, \textbf{36}, 113   \BIBCODE{1986AcA....36..113K}

Ka{\l}u\.{z}ny, J., 1986c, \textit{AcA}, \textbf{36}, 121   \BIBCODE{1986AcA....36..121K}

Ka{\l}u\.{z}ny, J., 1986d, \textit{PASP}, \textbf{98}, 662   \BIBCODE{1986PASP...98..662K}

Liakos, A., Niarchos, P., 2011, \textit{IBVS}, 5999, 2   \BIBCODE{2011IBVS.5999....5L} 

Lucy, L. B., 1967, \textit{ZsfAp}, \textbf{65}, 89   \BIBCODE{1967ZA.....65...89L}

Lucy, L. B., 1968, \textit{ApJ}, \textbf{151}, 1123  \BIBCODE{1968ApJ...151.1123L} \DOI{10.1086/149510}

Odell, A.P., Wils, P., Dirks, C., Guvenen, B., O'Malley, C.J., Villarreal, A.S., Weinzettle, R.M.,  2011, \textit{IBVS}, 6001   \BIBCODE{2011IBVS.6001....1O}

Popper, D. M., 1984, \textit{AJ}, \textbf{89}, 132   \BIBCODE{1984AJ.....89..132P} \DOI{10.1086/113491}

Rucinski, S.M., 1969, \textit{AcA}, \textbf{19}, 245   \BIBCODE{1969AcA....19..245R}

Samec, R. G., Kring, J. D., Robb, R., Van Hamme, W., Faulkner, D. R., 2015a, \textit{AJ}, \textbf{149}, 90  (\SAMEC)   \BIBCODE{2015AJ....149...90S} \DOI{10.1088/0004-6256/149/3/90}

Samec, R. G., Benkendorf, B., Dignan, J. B., Robb, R., Kring, J., Faulkner, D. R., 2015b, \textit{AJ}, \textbf{149}, 146   \BIBCODE{2015AJ....149..146S} \DOI{10.1088/0004-6256/149/4/146}

Siwak, M., Zola, S., Koziel-Wierzbowska, D., 2010, \textit{AcA}, \textbf{60}, 305   \BIBCODE{2010AcA....60..305S}

Van Hamme, W., 1993, \textit{AJ}, \textbf{106}, 2096   \BIBCODE{1993AJ....106.2096V} \DOI{10.1086/116788}

Wilson, R.E., Devinney, E.J., 1971,  \textit{ApJ}, \textbf{166}, 605   \BIBCODE{1971ApJ...166..605W} \DOI{10.1086/150986}

Wilson, R. E., 1990,  \textit{ApJ}, \textbf{356}, 613   \BIBCODE{1990ApJ...356..613W} \DOI{10.1086/168867}

Wilson, R. E., 1994,  \textit{PASP}, \textbf{106}, 921   \BIBCODE{1994PASP..106..921W} \DOI{10.1086/133464}

Wozniak, P. R. et al., 2004, \textit{AJ}, \textbf{127}, 2436 (NSVS)   \BIBCODE{2004AJ....127.2436W} \DOI{10.1086/382719}

\endreferences

\end{document}